\pdfoutput=1
\documentclass[aps,prl,twocolumn,floatfix,superscriptaddress]{revtex4}

\usepackage{amssymb,amsmath,amstext}                
\usepackage{graphicx}                                                
\usepackage{epstopdf}                                               
\usepackage{color}                                                     
\usepackage{bm}                                                        
\usepackage{appendix}                                              
\usepackage[utf8]{inputenc}
\usepackage{bbold}
\usepackage{bbm}
\usepackage{ulem}
\usepackage[shrink=41,letterspace=500]{microtype}
\usepackage{latexsym}
\usepackage[colorlinks=true,citecolor=blue,linkcolor=magenta]{hyperref}

\def\be{\begin{equation}}
\def\ee{\end{equation}}
\def\bea{\begin{eqnarray}}
\def\eea{\end{eqnarray}}

\def\bi{\begin{itemize}}
\def\ei{\end{itemize}}
\def\ben{\begin{enumerate}}
\def\een{\end{enumerate}}

\definecolor{ao}{rgb}{0., 0.5, 0.}

\definecolor{dgreen} {RGB}{78,138,21}

\begin{document} 

\title{Comment on ``Quantum Time Crystals and Interacting Gauge Theories in Atomic Bose-Einstein Condensates''}

\author{Andrzej Syrwid} 
\affiliation{
Instytut Fizyki Teoretycznej, 
Uniwersytet Jagiello\'nski, ulica Profesora Stanis\l{}awa \L{}ojasiewicza 11, PL-30-348 Krak\'ow, Poland}
\affiliation{Laboratoire Kastler Brossel, UPMC-Sorbonne Universités, CNRS, ENS-PSL Research University,Collège de France; 4 Place Jussieu, 75005 Paris, France
}
\author{Arkadiusz Kosior} 
\affiliation{
Instytut Fizyki Teoretycznej, 
Uniwersytet Jagiello\'nski, ulica Profesora Stanis\l{}awa \L{}ojasiewicza 11, PL-30-348 Krak\'ow, Poland}
\affiliation{Max-Planck-Institut f\"ur Physik Komplexer Systeme,
N\"othnitzer Strasse 38, D-01187, Dresden, Germany}
\author{Krzysztof Sacha} 
\affiliation{
Instytut Fizyki Teoretycznej, 
Uniwersytet Jagiello\'nski, ulica Profesora Stanis\l{}awa \L{}ojasiewicza 11, PL-30-348 Krak\'ow, Poland}


\date{\today}

\maketitle


In the recent Letter \cite{Ohberg2018}, \"Ohberg and Wright describe a Bose-Einstein condensate trapped on a ring in the presence of the density-dependent gauge potential. It is claimed that the ground state of the system corresponds to a rotating chiral bright soliton and consequently it forms a genuine time crystal which minimizes its energy by performing periodic motion. We show that the energy of the chiral soliton in the laboratory frame is not correctly calculated in the Letter. The correct energy becomes minimal if the soliton does not move.

The genuine time crystal would be a time-independent quantum system which spontaneously breaks the continuous time translation symmetry into a discrete time translation symmetry in its ground state \cite{Wilczek2012,Sacha2017rev}. In other words such a system spontaneously switches to periodic motion even if it has the lowest possible energy. Wilczek postulated that bosons with attractive interactions on the Aharonov-Bohm ring would form a bright soliton which performs periodic motion in the ground state \cite{Wilczek2012}.  However, it turned out that in the limit of a large number of bosons, the soliton does not move in the lowest energy state \cite{Bruno2013,Syrwid2017}. 
In the Letter \cite{Ohberg2018} a chiral bright soliton solution is analyzed and we show that, on the contrary to the claim of the authors, it also does not move if its energy is minimal.

In Ref.~\cite{Ohberg2018} the following energy per particle in the laboratory frame is considered:
\be
{\cal E}_{LAB}=\frac{1}{N}\int dx\; \Psi^*\left(\frac{(p-A)^2}{2m}+\frac{g}{2}|\Psi|^2\right)\Psi,
\label{elab}
\ee
where $A=-(\hbar/2) \partial_x \phi+a_1|\Psi|^2$ and we have chosen $W=0$, similarly like in the Supplemental Material of the Letter \cite{Ohberg2018}. The lowest energy solution shown in Ref.~\cite{Ohberg2018} is written in the following form:
\be
\Psi(x,t)=\sqrt{N}e^{-i\frac{\phi}{2}+\frac{ia_1N}{\hbar}\int^xds|\psi(s,t)|^2}\psi(x,t),
\label{sol1} 
\ee
where
\be
\psi(x,t) = e^{\frac{imux}{\hbar}-\frac{imu^2t}{2\hbar}}\Phi(x-ut,t). 
\label{sol2}
\ee
Let us simply substitute the above solution to the energy functional in Eq.~(\ref{elab}). The substitution in Eq.~(\ref{sol1}) leads to
\bea
{\cal E}_{LAB}&=&\int dx\;\psi^*\left(\frac{p^2}{2m}+\frac{gN}{2}|\psi|^2\right)\psi,
\eea
and employing the expression in Eq.~(\ref{sol2}) we get
\bea
{\cal E}_{LAB}&=&
\int dx\left[\frac{\hbar^2}{2m}\left|\frac{\partial\Phi}{\partial x}+\frac{imu}{\hbar}\Phi\right|^2+\frac{gN}{2}|\Phi|^4\right],
\label{elab1}
\eea
which is different from Eq.~(20) in the Letter \cite{Ohberg2018} and from Eq.~(14) in the Supplemental Material of the same Letter. 
That is, in comparison to Eq.~(\ref{elab1}),  instead of $g$ the authors have $\tilde g=(g-2a_1u)$ 
[where we have used the definition of $\tilde g$ given in the Letter before Eq.~(15)]

When we substitute in Eq.~(\ref{elab1}) the bright soliton solution considered in \cite{Ohberg2018} [see Eq.~(15) in the Letter], i.e. $\Phi(x-ut,t)=\chi(x-ut)e^{-i\mu t}$ where
\be\label{soliton}
\chi(s)=\frac{1}{\sqrt{2b}}\frac{1}{\cosh(s/b)},
\ee
with $b=-2\hbar^2/(m\tilde gN)$, we obtain
\be
{\cal E}_{LAB}=-\frac{mg^2N^2}{24\hbar^2}+\left(1+\frac{a_1^2N^2}{3\hbar^2}\right)\frac{mu^2}{2}.
\label{elab2}
\ee
Equation~(\ref{elab2}) indicates that in the lowest energy state the velocity of the soliton $u=0$ and no genuine time crystal behavior is observed.

In the Letter \cite{Ohberg2018}, the authors perform two time-dependent transformations every time calculating the energy in the corresponding reference frame by transforming the Lagrangian in the Dirac-Frenkel action. After the first transformation, the energy ${\cal E}$ in the new frame is given by Eq.~(9) of the Supplemental Material of the Letter. The second transformation to the moving frame leads to the energy ${\cal E}'$, Eq.~(13) of the Supplemental Material. In order to calculate the energy in the laboratory frame, the inverse transformations have to be performed. However, the authors do not return to the energy in the laboratory frame but to the energy $\cal E$. 

We do not perform the transformations of the energy to the different frames but simply substitute the solution, Eqs.~\eqref{sol1} and \eqref{sol2}, to the energy functional in the laboratory frame Eq.~(\ref{elab}).


To conclude, the genuine time crystal does not form in the system considered in Ref.~\cite{Ohberg2018}. In the thermodynamic limit, systems with two-body interactions cannot form time crystals in the equilibrium state \cite{Watanabe2015,Watanabe2019}.

Support of the National Science Centre, Poland via Projects No.~2018/28/T/ST2/00372 (A.S.), No.~2016/21/B/ST2/01086 (A.K.) and No.~2018/31/B/ST2/00349
(K.S) is acknowledged. A.S. and A.K. acknowledge the support of the Foundation for Polish Science (FNP). 


\bibliography{ref_tc_book}

\begin{thebibliography}{7}
\expandafter\ifx\csname natexlab\endcsname\relax\def\natexlab#1{#1}\fi
\expandafter\ifx\csname bibnamefont\endcsname\relax
  \def\bibnamefont#1{#1}\fi
\expandafter\ifx\csname bibfnamefont\endcsname\relax
  \def\bibfnamefont#1{#1}\fi
\expandafter\ifx\csname citenamefont\endcsname\relax
  \def\citenamefont#1{#1}\fi
\expandafter\ifx\csname url\endcsname\relax
  \def\url#1{\texttt{#1}}\fi
\expandafter\ifx\csname urlprefix\endcsname\relax\def\urlprefix{URL }\fi
\providecommand{\bibinfo}[2]{#2}
\providecommand{\eprint}[2][]{\url{#2}}

\bibitem[{\citenamefont{\"Ohberg and Wright}(2019)}]{Ohberg2018}
\bibinfo{author}{\bibfnamefont{P.}~\bibnamefont{\"Ohberg}} \bibnamefont{and}
  \bibinfo{author}{\bibfnamefont{E.~M.} \bibnamefont{Wright}},
  \bibinfo{journal}{Phys. Rev. Lett.} \textbf{\bibinfo{volume}{123}},
  \bibinfo{pages}{250402} (\bibinfo{year}{2019}),
  \urlprefix\url{https://link.aps.org/doi/10.1103/PhysRevLett.123.250402}.

\bibitem[{\citenamefont{Wilczek}(2012)}]{Wilczek2012}
\bibinfo{author}{\bibfnamefont{F.}~\bibnamefont{Wilczek}},
  \bibinfo{journal}{Phys. Rev. Lett.} \textbf{\bibinfo{volume}{109}},
  \bibinfo{pages}{160401} (\bibinfo{year}{2012}),
  \urlprefix\url{http://link.aps.org/doi/10.1103/PhysRevLett.109.160401}.

\bibitem[{\citenamefont{{Sacha} and {Zakrzewski}}(2018)}]{Sacha2017rev}
\bibinfo{author}{\bibfnamefont{K.}~\bibnamefont{{Sacha}}} \bibnamefont{and}
  \bibinfo{author}{\bibfnamefont{J.}~\bibnamefont{{Zakrzewski}}},
  \bibinfo{journal}{Rep. Prog. Phys.} \textbf{\bibinfo{volume}{81}},
  \bibinfo{pages}{016401} (\bibinfo{year}{2018}),
  \urlprefix\url{https://doi.org/10.1088/1361-6633/aa8b38}.

\bibitem[{\citenamefont{Bruno}(2013)}]{Bruno2013}
\bibinfo{author}{\bibfnamefont{P.}~\bibnamefont{Bruno}},
  \bibinfo{journal}{Phys. Rev. Lett.} \textbf{\bibinfo{volume}{110}},
  \bibinfo{pages}{118901} (\bibinfo{year}{2013}),
  \urlprefix\url{http://link.aps.org/doi/10.1103/PhysRevLett.110.118901}.

\bibitem[{\citenamefont{Syrwid et~al.}(2017)\citenamefont{Syrwid, Zakrzewski,
  and Sacha}}]{Syrwid2017}
\bibinfo{author}{\bibfnamefont{A.}~\bibnamefont{Syrwid}},
  \bibinfo{author}{\bibfnamefont{J.}~\bibnamefont{Zakrzewski}},
  \bibnamefont{and} \bibinfo{author}{\bibfnamefont{K.}~\bibnamefont{Sacha}},
  \bibinfo{journal}{Phys. Rev. Lett.} \textbf{\bibinfo{volume}{119}},
  \bibinfo{pages}{250602} (\bibinfo{year}{2017}),
  \urlprefix\url{https://link.aps.org/doi/10.1103/PhysRevLett.119.250602}.

\bibitem[{\citenamefont{Watanabe and Oshikawa}(2015)}]{Watanabe2015}
\bibinfo{author}{\bibfnamefont{H.}~\bibnamefont{Watanabe}} \bibnamefont{and}
  \bibinfo{author}{\bibfnamefont{M.}~\bibnamefont{Oshikawa}},
  \bibinfo{journal}{Phys. Rev. Lett.} \textbf{\bibinfo{volume}{114}},
  \bibinfo{pages}{251603} (\bibinfo{year}{2015}),
  \urlprefix\url{http://link.aps.org/doi/10.1103/PhysRevLett.114.251603}.

\bibitem[{\citenamefont{Watanabe et~al.}(2019)\citenamefont{Watanabe, Oshikawa,
  and Koma}}]{Watanabe2019}
\bibinfo{author}{\bibfnamefont{H.}~\bibnamefont{Watanabe}},
  \bibinfo{author}{\bibfnamefont{M.}~\bibnamefont{Oshikawa}}, \bibnamefont{and}
  \bibinfo{author}{\bibfnamefont{T.}~\bibnamefont{Koma}},
  \emph{\bibinfo{title}{Proof of the absence of long-range temporal orders in
  gibbs states}} (\bibinfo{year}{2019}), \eprint{1911.12939}.

\end{thebibliography}


\end{document}